\documentclass{aastex631}
\usepackage{multirow}
\shorttitle{3D Hydrodynamical Simulations of He-ignited DD WD Mergers}
\graphicspath{{./}{figures/}}

\begin{document}

\title{3D Hydrodynamical Simulations of Helium-Ignited Double-degenerate White Dwarf Mergers}

\correspondingauthor{Robert Fisher}
\email{robert.fisher@umassd.edu}

\author[0000-0002-0487-3090]{Niranjan C. Roy}
\affiliation{Department of Physics, University of Massachusetts Dartmouth,\\
285 Old Westport Road,\\
North Dartmouth, MA 02740, USA}
\affiliation{Department of Physics, University of Connecticut,\\ Storrs, CT 06269, USA}
\nocollaboration{1}

\author[0000-0002-7110-9885]{Vishal Tiwari}
\affiliation{Department of Physics, University of Massachusetts Dartmouth,\\
285 Old Westport Road,\\
North Dartmouth, MA 02740, USA}
\affiliation{Center for Relativistic Astrophysics, School of Physics, Georgia Institute of Technology,\\ Atlanta, GA 30332, USA}
\nocollaboration{1}

\author[0000-0002-4674-0704]{Alexey Bobrick}
\affiliation{Technion - Israel Institute of Technology, Physics Department,\\ Haifa, Israel 32000}
\affiliation{Lund Observatory, Department of Astronomy and Theoretical physics, Box 43,\\ SE 221-00 Lund, Sweden}
\nocollaboration{1}

\author[0000-0002-9275-4584]{Daniel Kosakowski}
\affiliation{Department of Physics, University of Massachusetts Dartmouth,\\
285 Old Westport Road,\\
North Dartmouth, MA 02740, USA}
\nocollaboration{1}

\author[0000-0001-8077-7255]{Robert Fisher}
\affiliation{Department of Physics, University of Massachusetts Dartmouth,\\
285 Old Westport Road,\\
North Dartmouth, MA 02740, USA}
\nocollaboration{1}

\author[0000-0002-5004-199X]{Hagai B. Perets}
\affiliation{Technion - Israel Institute of Technology, Physics Department,\\ Haifa, Israel 32000}
\nocollaboration{1}

\author[0000-0002-5700-282X]{Rahul Kashyap}
\affiliation{Institute for Gravitation and the Cosmos, The Pennsylvania State University,\\University Park, PA 16802, USA}
\affiliation{Department of Physics, The Pennsylvania State University,\\ University Park, PA 16802, USA} 
\affiliation{ Department of Astronomy and Astrophysics, The Pennsylvania State University,\\ University Park, PA 16802, USA}
\nocollaboration{1}

\author[0000-0001-8627-6478]{Pablo Lorén-Aguilar}
\affiliation{5 School of Physics, University of Exeter, Stocker Road, Exeter EX4 4QL, UK}
\nocollaboration{1}

\author[0000-0002-1623-5838]{Enrique García-Berro}
\affiliation{Departament de Física Aplicada, Universitat Politècnica de Catalunya, c/Esteve Terrades, 5, E-08860 Castelldefels, Spain}
\nocollaboration{1}

\begin{abstract}

The origins of type Ia supernovae (SNe Ia) are still debated. Some of the leading scenarios involve a double detonation in double white dwarf (WD) systems. In these scenarios, helium shell detonation occurs on top of a carbon-oxygen (CO) WD, which then drives the detonation of the CO-core, producing a SN Ia. Extensive studies have been done on the possibility of a double helium detonation, following a dynamical helium mass-transfer phase onto a CO-WD. However, 3D self-consistent modeling of the double-WD system, the mass transfer, and the helium shell detonation have been little studied. 
Here we use 3D hydrodynamical simulations to explore this case in which a helium detonation occurs near the point of Roche lobe overflow of the donor WD and may lead to an SN Ia through the dynamically driven double-degenerate double-detonation (D6) mechanism. 
We find that the helium layer of the accreting primary WD does undergo a detonation, while the underlying carbon-oxygen core does not, leading to an extremely rapid and faint nova-like transient instead of a luminous SN Ia event. This failed core detonation
suggests that D6 SNe Ia may be restricted to the most  massive carbon-oxygen primary WDs. We highlight the nucleosynthesis of the long-lived radioisotope $^{44}$Ti during explosive helium burning, which may serve as a hallmark both of successful as well as failed D6 events which subsequently detonate as classical double-degenerate mergers. 
\end{abstract}

\section{Introduction}\label{sec:intro}
Type Ia supernovae (SNe Ia) are important tools in our understanding of the physical universe, producing most of the iron elements in the universe \citep{jha2019},
and serving  as standardizable candles to gauge cosmological distances \citep{Perlmutter_1999, riess..1998AJ....116.1009R}. Constraining the SNe Ia stellar progenitors  has been a primary focus of research for the last few decades. 
 The discovery of three hypervelocity WDs in the the Gaia DR2 catalog provided potential observational evidence of the ex-companions of  sub-$M_{\rm Ch}$ WDs which underwent SNe Ia explosions in a variant of the double-degenerate scenario of binary white dwarf mergers \citep{shen2018three}. 
In this proposed variant of the double-degenerate scenario, dubbed the dynamically driven double-degenerate double-detonation, or D6, scenario, two CO WDs with thin surface helium layers remaining from single stellar evolution can lead to a SN Ia   \citep{shen2018sub}.
In the D6 scenario, the convergence of the helium detonation front on the accretor causes a second detonation inside the accretor's CO core. It is this second detonation in the CO core which completely disrupts the accretor, and gives rise to a type Ia event.\\ 

To date, only a handful of simulations have investigated the role of thin helium layers in double degenerate mergers. \citet {Guillochon_etal_2010}, \citet {tanikawa2019double}, and \citet{pakmor2022} explored 
three-dimensional adaptive mesh and smoothed particle hydrodynamical (SPH) simulations of merging with sub-$M_{\rm Ch}$ WDs containing helium envelopes.
\citet{pakmor2021} 
have studied dynamical mass-transfer of helium from a hybrid HeCO WD \cite[see][for discussion of Ia SNe from hybrid HeCO disruptions]{Zen+18,PeretsZenati20} onto a 0.8 M$_\odot$ CO WD (with no initial helium envelope) using the 3D moving-mesh code AREPO. \citet {pakmor2022} demonstrated a successful D6 scenario with a 1.05 M$_{\odot}$ primary, a 0.7 M$_{\odot}$ secondary WD, and a total helium mass of 0.06 M$_{\odot}$; all other models have not led to the detonation of the CO WD primary. \\

In this paper, we address several key questions concerning the D6 scenario. Specifically, how does the surface helium detonation develop under realistic conditions including the full three-dimensionality of the accretion stream and rotation? Under what conditions does the D6 scenario detonate the accretor core, and lead to a successful SN Ia event? Under what conditions does the scenario fail to detonate the accretor core? When the scenario ignites only the helium surface layer on the accretor, what are the signatures of the transient, its nucleosynthetic products, and what are the implications for the subsequent merger of the two CO WDs?\\

In order to address these key science questions, we perform three-dimensional hydrodynamical simulations of the full binary WD system, with a massive primary CO (1 M$_\odot$) WD and a typical 0.6 M$_{\odot}$ mass CO WD donor. Both WDs have  thin surface helium layers. The full three-dimensionality ensures that the geometry of the accretion stream is fully captured, and that artificial symmetries imposed in two dimensional and three dimensional non-rotating simulations are broken. We present our work with the simulation setup and convergence study in section \ref{sec: setup}, results in section \ref{sec:results}, and conclusions and discussion in section \ref{conclusion}.\\

\section{Simulation setup}
\label{sec: setup}
We explore the role of helium in double-degenerate mergers, choosing a 1 M$_{\odot}$ CO WD accretor and 0.6 M$_{\odot}$ CO WD donor. Both WDs have a carbon to oxygen mass ratio of $\sim0.67$, and each also begins with 0.01 M$_{\odot}$ of helium on their surfaces, and $\sim$0.01 M$_{\odot}$ of helium in the accretion stream. We have chosen the mass of helium layer on the primary to be higher than the lower limit needed for a detonation in a double-degenerate system \citep{kato2008helium}. The helium masses are within the upper limit of the  critical mass of 0.05 M$_{\odot} $\citep{ruiter2014heliumdet}; thicker helium layers ($> 0.05 {\rm M}_{\odot}$) in the WDs will have helium ash products  produced after the detonation which is in tension with the observed spectra \citep{2011ApJ...734...38W} and will not be considered here. It is also not clear whether such thicker helium layers can survive on massive WDs without nuclear burning beforehand \citep{ShenBildsten2009ApJ...699.1365S}.\\

Our hydrodynamical simulations take advantage of both the mesh-free Lagrangian and grid-based Eulerian methods. Our initial conditions are generated by building upon the strengths of each method. Specifically, the SPH method is able to accurately model the inspiral phase of the merger and the initial stages of mass accretion over many orbits, while {\fontfamily{qcr}\selectfont
FLASH} AMR grid captures hydrodynamical instabilities and detonation fronts accurately. The SPH simulation \citep{Loren_Aguilar_2010, bobrick2017mass}  of the double-degenerate merger follows the WD binaries using the Helmholtz equation of state \citep{Timmes_2000} through a point where dynamical burning becomes important. The SPH data is then mapped to the {\fontfamily{qcr}\selectfont
FLASH}  Eulerian adaptive mesh refinement (AMR) code \citep{fryxell2000flash} at a time, based on the mass of helium burnt. In the following, we reckon $t = 0$ relative to this remapping time.\\

We use the scatter algorithm to initialize the {\fontfamily{qcr}\selectfont
FLASH} grid from the SPH data. This method involves the smoothing of properties of the particles at a particular cell on the grid by adding the weighted contributions of all the particles.
We use the cubic piecewise smoothing kernel given in equation (\ref{smkernel}) to map the SPH hydrodynamical primitive fields to the AMR grid \citep{hernquist1989treesph}.\\

\begin{equation}
\label{smkernel}
W(|\vec{r} - \vec{r_j}|,h_j) = \left\{
        \begin{array}{ll}
      \frac{1}{\pi h_j^3} \left\{ 1 - \frac{3}{2} \left(\frac{|\vec{r} - \vec{r_j}|}{h_j}\right)^2 + \frac{3}{4} \left(\frac{|\vec{r} - \vec{r_j}|}{h_j}\right)^3\right\} & 0 \leq \frac{|\vec{r} - \vec{r_j}|}{h_j} \leq 1 \\

      \frac{1}{\pi h_j^3} \left\{\frac{1}{4}\left[2 -\left(\frac{|\vec{r} - \vec{r_j}|}{h_j}\right)\right]^3\right\} & 1 < \frac{|\vec{r} - \vec{r_j}|}{h_j} \leq 2 \\
      0 & \frac{|\vec{r} - \vec{r_j}|}{h_j} > 2
        \end{array}
    \right.
\end{equation}

\begin{equation}
\label{smproperty}
  A(\vec{r}) = \sum_{j=0}^{N}m_j \frac{A_j}{\rho_j} W(|\vec{r} - \vec{r_j}| , h_j)
\end{equation}
\vspace{2mm}

Here, $h_j$ is the smoothing length of the $j$th particle at $\vec {r_j}$, and $A_j$ represents any of the primitive hydrodynamic fields (density, velocities, and temperature) of the $j$th particle being mapped to the AMR cell centered at location $\vec {r}$. After mapping the hydrodynamic primitive fields to the cells and converting to the conservative hydrodynamic basis, the calculation of the internal energy and the pressure on the AMR mesh is then performed using the same Helmholtz equation of state used by both codes.\\

The 3D AMR computational domain is set to be $5.6\times10^{5}$ km in length in each spatial dimension, with diode boundary conditions on all sides.
 We employ a mass-based resolution criterion, ensuring that no cell exceeds  $5\times10^{26}$ g. A temperature based refinement is also employed with a threshold value of $10^9$ K. In both cases, after a refinement trigger is reached within a cell, the whole block  is refined.
We use the directionally unsplit hydro solver available in {\fontfamily{qcr}\selectfont
FLASH} for hydrodynamics and a modified multipole gravity solver \citep{Couch_2013} up to $l=60$ for the self-gravity calculations, with isolated boundary conditions.   We undertook a series of four simulations of increasing finest grid resolution, from 136 km to 17 km. 
To capture the nuclear energetics of the burning, we incorporate the 19 isotope alpha chain nuclear network Approx19 \citep{Timmes_1999} with our hydrodynamical simulations.\\

\section{Results}
\label{sec:results}

In this section we present 3D hydrodynamical simulations of the double-degenerate merger.

\subsection{Helium detonation}
 As the accretion stream transports mass from the donor to the accretor, Kelvin-Helmholtz instabilities occur along the interface of the accretion stream with the outer layers of the accretor. When the eddy turnover timescale in such a turbulent environment becomes comparable to the local burning timescale, the turbulent heating may give rise to a detonation  \citep{shen2014initiation, 2019ApJ...876...64F}. The detonation front initiated in the helium surface layers of the accretor then propagates, both as a detonation front wrapping around the helium layer, and also as a non-burning shock front through the CO core. \\

\begin{figure}[bp!]
    \centering
    \includegraphics[width=1\textwidth]{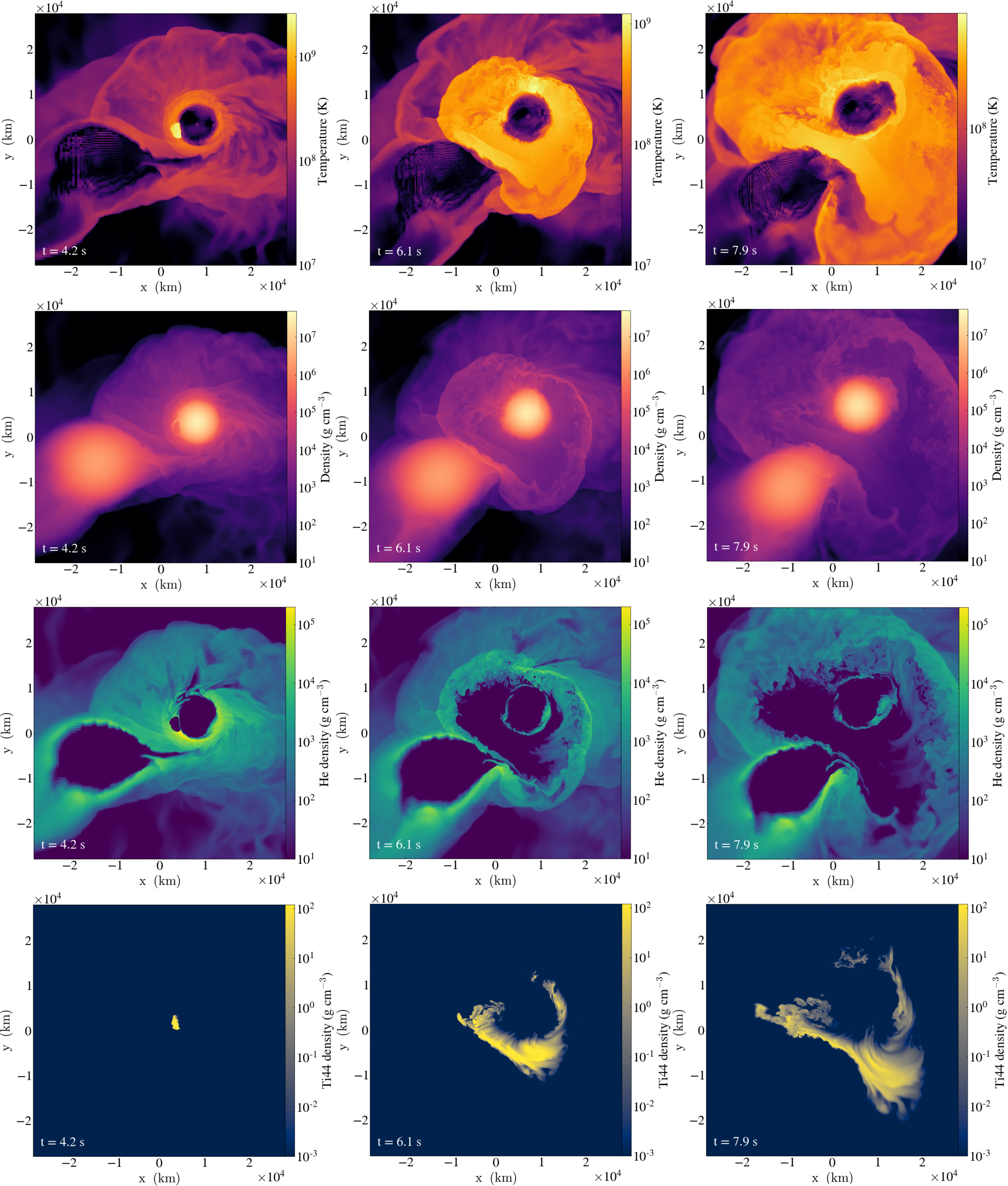}
    \caption{Evolution of key hydrodynamic and nucleosynthetic quantities through the $Z = 0$ midplane during and after the helium detonation on the accretor. From top to bottom, the temperature, mass density, $^4$He, and $^{44}$Ti mass densities are shown at times $t = (4.2, 6.1, 7.9)$ s, at a resolution of 17 km.}
    \label{fig:profiles_together}
\end{figure}

 In Figure \ref{fig:profiles_together}, key physical quantities of the simulation are presented as slice plots through the $z = 0$ midplane  at three different moments ($t = 4.3, 6.3, 7.9$ s). In the left column of panels, the temperature, density, $^4$He density and the $^{44}$Ti density profiles are shown at the time when detonation starts in the helium layer. The detonation fronts then propagate over the surface of the accretor and collide, but fail to detonate the underlying CO core, as shown in the middle column of the figure. The expanding detonation front is strong enough to affect the dynamics of the accretion stream.
 The rightmost plots depict a time when the detonation front has swept over the accretion stream; temporarily disrupting the accretion, and ablating helium from  the donor's surface. We note that in these simulations, the expanding helium front does not ignite the surface helium on the secondary.\\ 
 
\begin{figure}
    \centering
    \includegraphics[width=1\textwidth]{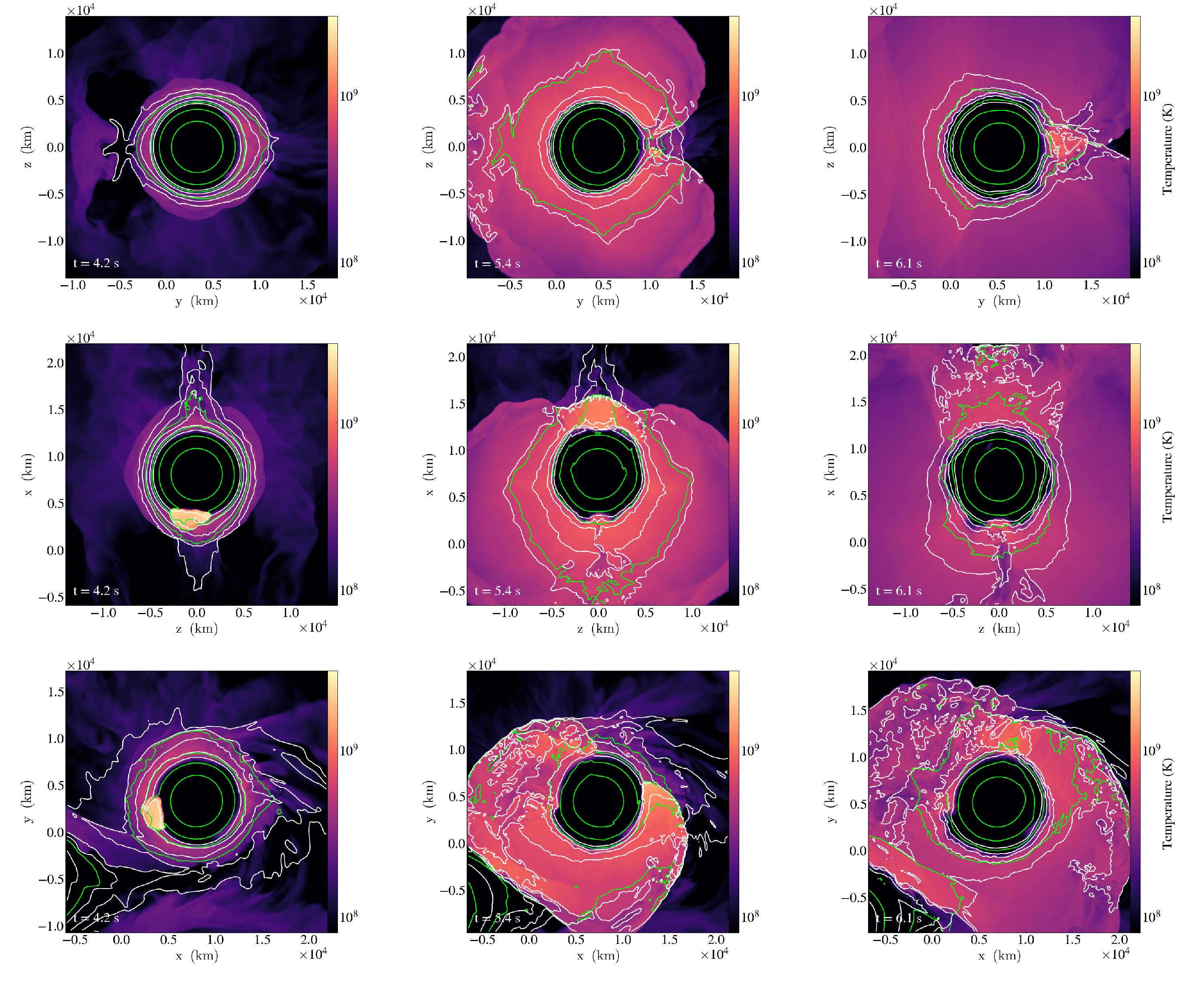}
     \caption{Density (white) and pressure (lime) contours overlaid on 2D temperature slice plots along the $X = 0$ (top row) midplane, and corresponding planes along $Y$ (middle row) and $Z$ (bottom row) at times $(4.2, 5.4, 6.1)$ s for a resolution of 17 km.
    \label{fig:DenPresContour}}
\end{figure}

 Another key feature of the helium density profile on the accretor is its asymmetry. The helium envelope is oblate, owing to rotation. The departure from spherical symmetry is most evident in the outer envelope along cross slices through  $X$ and $Y$ midplane, visible in Figure \ref{fig:DenPresContour} (first entry of top row and first entry of middle row). We see that the outer density contours are indeed stretched along the $Y$ axis, with the innermost density contours approaching spherical symmetry.
Further, looking at the position of the helium detonation initiation, it is also evident that the runaway nuclear reaction in this layer starts from a point which is within this accretion plane.\\

Soon after the burning and ablation phase has taken place, the accretion stream from the accretor is revived, and again helium begins to accumulate on the surface of the primary. Such a revival of accretion may result in a second helium surface detonation,  before the helium budget of the system is completely exhausted. The outcome could either be a successful CO core detonation, or the disruption of the the donor and the complete merger of the system, similar to a canonical double-degenerate merger. \\
 
\begin{figure}[bp!]
    \centering
    \includegraphics[width = 0.6\textwidth]{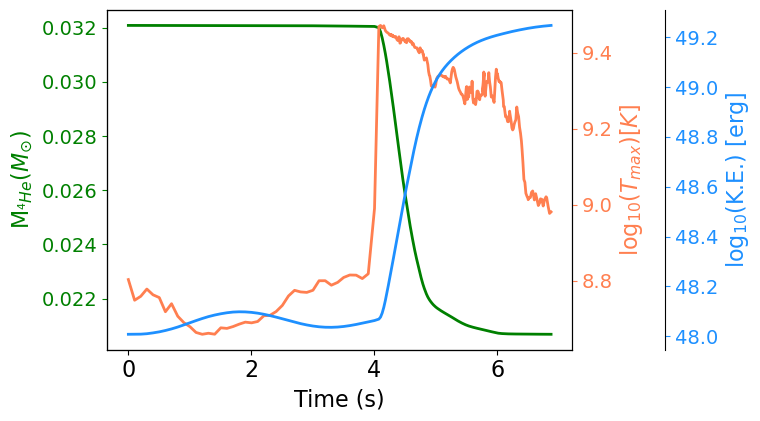}
    \caption{ Evolution of global maximum temperature, kinetic energy (excluding rotational kinetic energy), and total $^4$He budget in the computational domain. This plot shows the initial helium detonation at $t = 4.1$ s and the subsequent collision of detonation fronts at $t \sim 6$ s.
    \label{fig:He4MaxTempEvolution}}
   
\end{figure}

While the helium surface burning on the primary produces no significant amount of $^{56}$Ni, the burned material is relatively abundant in intermediate mass elements, including $^{28}$Si, $^{40}$Ca, as well as the radioisotope $^{44}$Ti (Table \ref{tab:NucleoYields}). $^{44}$Ti is a key radioisotope owing to its long half life of 60 yr, and which has been directly detected through gamma rays in galactic SNRs Tycho and Cas A \citep {weinbergeretal20}. The mass density of $^{44}$Ti is shown in the bottom row of panels in Figure \ref{fig:profiles_together}.  It is evident that the $^{44}$Ti is nucleosynthesized during surface helium burning on the primary, as earlier simulations have found in the context of the double-detonation scenario \citep {woosleytaamweaver86}.
The detonation initiation is seen in the 17 km resolution run in the temperature slice plot in the top-left corner of Figure \ref{fig:profiles_together}, taken  through the $Z=0$ midplane $t = 4.2$ s after the start of the FLASH simulation. The detonation fronts then propagate and collide when they reach a nearly opposite point on the accretor, giving rise to a second high temperature peak after the detonation at $t = 6.1$ s. These times are readily identifiable from the time history of the global helium mass, shown in figure \ref{fig:He4MaxTempEvolution}.\\

Figure \ref{fig:He4MaxTempEvolution} also shows the evolution of kinetic energy, excluding the rotational energy, as a global quantity in the simulation domain.
The global kinetic energy increases with time after helium detonation initiation, and reaches its peak shortly after the detonation front collision.
If the high temperatures in the shock collision region occur in CO-rich region of the accretor,  a runaway nuclear reaction would be produced in the CO core, causing a SN Ia.
However, in our simulations, the peak temperatures reached in the CO core remain well below the critical temperatures required for detonation initiation, and the CO core does not undergo a detonation.\\ 

An important result unique to the fully three-dimensional simulations of the binary presented here, which include rotation, is the geometry of the helium detonation front and the collision region. To constrain the geometry of the collision region, we overlay the density and pressure contours on top of 2D slice plot of temperature on the accretor through three different planes (Figure \ref{fig:DenPresContour}). The upper and lower limits of the five density contours have been selected as $5\times10^3$ g cm$^{-3}$ and $5\times10^5$ g cm$^{-3}$, and that of the pressure has been chosen as $5\times10^{20}$ dyne cm$^{-2}$ and $5\times10^{24}$ dyne cm$^{-2}$ respectively. 
The location of this convergence is approximately 8500 km away from the center of the accretor. The density, pressure and temperature at the point of shock convergence are found to be $5.48 \times 10^5$ g/cm$^3$, $1.47\times10^{23}$ dyne/cm$^2$, and $2.30\times10^9$K respectively, in the 17 km run. Just before the onset of detonation, the $^{12}$C, $^{16}$O, and $^4$He mass fractions at the point of ignition were 0.22, 0.33, and 0.45 respectively, indicating the initiation location had an admixture of helium and underlying CO. \citet{shen2014initiation} suggest such a pollution of CO in the helium layer may enhance the possibility of a helium detonation.

\subsection{Convergence study of the helium detonation}
We present here the convergence of the helium detonation in terms of increasing resolutions of 136 km, 68 km, 34 km, and 17 km. The temperature at the location of helium detonation initiation ranges from 2.38$\times 10^9$ K to 2.93$\times 10^9$ K with maximum resolutions of the domain ranging from 136 km to 17 km respectively.
  As the resolution is increased, the maximum temperature approaches convergence. Specifically, the maximum temperature increases by 21.4\% from 136 km to 68 km. The fractional temperature deviation becomes 0.6\% with the subsequent increase of resolution to 34 km and 17km, which is a clear indication of the convergence.  Next in order, the temperature at the point of shock front convergence ranges from 2.03$
\times 10^9$ K to 2.30$\times 10^9$ K with the above mentioned range of resolutions. However, owing to the rotationally-induced broken symmetry of the system, even with the highest resolution of 17 km, this temperature is not enough to trigger a second detonation in the CO core.\\

\subsection{Nucleosynthetic abundances}

\begin{figure}[ht!]
    \centering
    \includegraphics[width=0.5\textwidth]{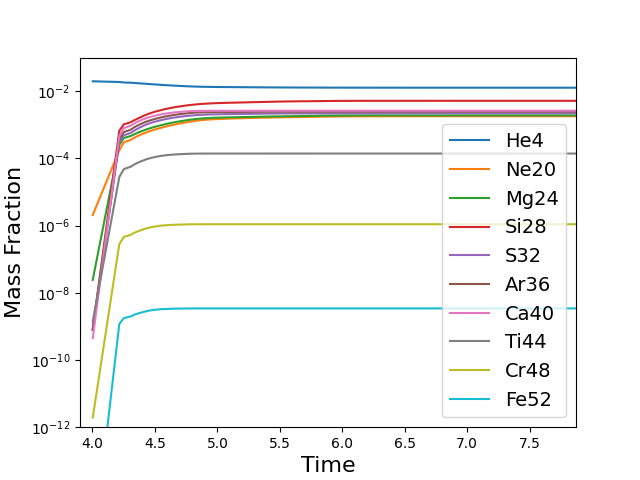}
    \caption{Temporal evolution of nuclosynthetic abundances.}
    \label{fig:SpeciesEvolution}
\end{figure}

Figure \ref{fig:SpeciesEvolution} depicts the evolution of the global nucleosynthetic abundances of the key species after detonation initiation, over the domain throughout time.
The mass fractions of $^{12}$C and $^{16}$O have not been included in this figure as they are dominated by the unburned donor and accretor, and consequently remain close to constant global values. The total amount of helium  at the beginning of simulation is 0.032 M$_{\odot}$.
After  nuclear burning, the amount of helium remaining becomes approximately 0.021 M$_{\odot}$. Consequently, the total helium budget of the system decreased by roughly 34\% over this cycle of mass accretion and burning.

At the onset of the helium detonation, there is a sharp increase in intermediate mass products such as $^{20}$Ne and $^{28}$Si, but without a subsequent increase of $^{56}$Ni.
The burning pollutes the accretor with intermediate mass elements, including $^{20}$Ne, $^{28}$Si, and $^{32}$S. The helium explosion ash on the primary can potentially impact the energetics of a possible second helium detonation upon the revival of the accretion stream, owing to the increased abundance of alpha capture heavy nuclei within the surface layers. \\

\begin{table}[ht!]
\centering
\begin{tabular}{l|l|l}
\toprule
 Species & Bound Mass (M$_{\odot}$) & Unbound Mass (M$_{\odot}$)\\
 \hline
 $^4$He & 1.6 $\times10^{-2}$ & $4.68\times 10^{-3}$ \\
 $^{12}$C & 6.32 $\times10^{-1}$ & $3.371 \times 10^{-3}$ \\
 $^{16}$O  & 9.41 $\times 10^{-1}$ & $1.515\times 10^{-3}$ \\
 $^{20}$Ne&  2.572 $\times 10^{-3}$ & $3.473 \times 10^{-4}$  \\
 $^{24}$Mg &  2.58 $\times 10^{-3}$ & $5.17 \times 10^{-4}$  \\
 $^{28}$Si &  5.73 $\times 10^{-3}$ & $2.693  \times 10^{-3}$ \\
 $^{32}$S &  2.144 $\times 10^{-3}$ & $1.435 \times 10^{-3}$   \\
 $^{36}$Ar & 2.011 $ \times 10^{-3}$ & $1.988  \times 10^{-3}$   \\
 $^{40}$Ca  & 1.574 $ \times 10^{-3}$ & $2.698 \times 10^{-3}$\\
 $^{44}$Ti  & 7.02 $\times 10^{-5}$  & $1.567\times 10^{-4}$ \\
 $^{48}$Cr & 5.24 $\times 10^{-7}$ & $1.262\times 10^{-6}$ \\
 $^{52}$Fe & 1.553 $\times 10^{-9}$ & $4.03\times 10^{-9}$  \\
\hline
\hline
\end{tabular}
\caption{Bound and unbound nucleosynthetic yields of key species. The final total bound mass is 1.606 $M_{\odot}$ and unbound mass 1.94 $\times 10^{-2} M_{\odot}$. Species with masses lower than $10^{-9} M_{\odot}$ are not included here.}
\label{tab:NucleoYields}
\end{table}

\section{conclusion}
\label{conclusion}

Our simulations of helium-ignited WD mergers exhibit a primary helium detonation, but do not immediately achieve a double-detonation in the accretor. The first helium detonation fails to create core detonation conditions even at spatial resolutions as high as 17 km. The direct outcome of this model will therefore be a faint helium nova, observable at galactic distances. \citet{kenshen2010pointIa} proposed that such a dynamical helium detonation event would produce a faint transient called ``.Ia'', but ``.Ia'' events are far more luminous than those expected in our models, where a very small amount of He is burnt, and very little radioactive materials are produced. The resulting outburst might be comparable to helium novae, like V445 Puppis  \citep{kato2003heliumnovae, goranskij2010progenitor}.
Consecutive helium detonations might also take place as a result of accretion before the donor fully disrupts. The system could possibly then give rise to a recurrent helium nova over the accretion timescale of the helium stream.\\

A caveat of these simulations is the relatively modest size of the 19 isotope nuclear network, which suffices for carbon and oxygen burning, but may have limitations for helium burning. It has been suggested by some authors that  larger (55+ isotopes) networks may better capture the energetics of helium burning \citep{townsley2019doubledet} during double detonations. However, even with the 19 isotope network, our models have shown that helium detonation occurs. If a larger network were to produce a stronger detonation, it might assist in enhancing the converging shock front into the CO core, but it might also produce an earlier helium detonation, with less accreted material available. So it is unclear what effect a larger nuclear network will have upon the CO core detonation. We do note that \citep{pakmor2021} used a larger network and found comparable energetics as we found here when exploring a similar-mass CO WD accretor.\\

\begin{acknowledgments}
The WDs in our simulations survive the initial phase of helium burning, and will continue to accrete carbon, oxygen, and helium  from the donor. This accretion process can end with a total disruption of the donor and a classic double-degenerate merger. The majority of the helium will have burned, and only traces will remain. We can not rule out that a fraction of these mergers then subsequently detonate and produce a normal SNe Ia. Consequently, a failed core detonation within a majority of binary white dwarf mergers may  account for the relatively few  hypervelocity WDs found in Gaia catalogue, in comparison to the much larger numbers expected if the D6 channel were to account for all normal SNe Ia. Alternatively, other scenarios such as a complete detonation of the donor immediately after the detonation of the primary may provide an explanation for the observed number of hypervelocity WDs \citep{pakmor2021}. Future observational tests should be able to discriminate between these two outcomes. 
\end{acknowledgments}

Our simulations demonstrate that $^{44}$Ti is produced during explosive helium burning in white dwarf mergers in yields comparable to those of core collapse supernovae remnants (SNRs) \citep{weinbergeretal20}. In the simulations presented here, the helium is burned in the minutes before the final merger of the white dwarfs, which could give rise to a SN Ia through a classical double degenerate merger scenario if the continued helium accretion fails to detonate the primary CO core. In another recent study  \citep{pakmor2022}, the surface helium yields a successful SN Ia. {\it Crucially, because this phase of surface helium burning must occur in any white dwarf merger, the $^{44}$Ti produced during explosive helium burning is a hallmark of the double-degenerate channel itself}. Specifically, the double-degenerate channel must necessarily involve the merger of two sub-$M_{\rm Ch}$ WDs with some surface helium on the primary and the secondary. The amount of $^{44}$Ti produced in explosive helium burning in both our simulations and those of \citet {pakmor2022} span the range of $\sim 10^{-4}$ - $10^{-3} {\rm M}_{\odot}$, nearly one to two orders of magnitude greater than the yields found during alpha-rich freezeout in near-$M_{\rm Ch}$ WDs \citep{leung2018explosive}. Thus, $^{44}$Ti is a crucial diagnostic which separates the broad class of double degenerate mergers of sub-$M_{\rm Ch}$ WDs, whether by a D6 or a classical merger, from the canonical single-degenerate near-$M_{\rm Ch}$ SN Ia scenario.\\

The $^{44}$Ti produced is observable in $\gamma$-rays for galactic SNe Ia and SNRs, as well as in the late-time optical light curves of SNe Ia in nearby galaxies, potentially long afterwards any associated SN Ia, due to its 60 yr half life \citep{The2006, woosley1986ApJ...301..601W}.
Of the SNe Ia SNRs, only SNR 1572 (Tycho) has a detection of $^{44}$Ti; Kepler and G1.9+0.3 have upper bounds only \citep{Weinberger2020A&A...638A..83W}.
INTEGRAL/IBIS 
inferred a $^{44}$Ti mass for Tycho of $1.5\pm1\times10^{-4}$ M$_{\odot}$ \citep{Renaud2006}, consistent with our calculated $^{44}$Ti yield. Additionally, the INTEGRAL/IBIS upper bounds of $^{44}$Ti mass of SNe Ia SNRs Kepler and G1.9+0.3 are 4$\times 10^{-4}$ M$_{\odot}$ and 0.3$\times 10^{-4}$ M$_{\odot}$, respectively \citep{Weinberger2020A&A...638A..83W}. Our calculated $^{44}$Ti yield is within the upper bound for Kepler, but a factor of 3 higher than the upper bound for the young remnant G1.9+0.3.\\

An even larger amount of $^{44}$Ti, as inferred by  \citet {pakmor2022}, would herald a D6-like event; the range of inferred gamma ray fluxes spanning both sets of helium-ignited WD merger simulations for a galactic event at a distance of 10 kpc is  $\sim \times 10^{-4} - 10^{-3}\ \gamma$ cm$^{-2}$s$^{-1}$, much larger than the measured flux from Cas A several centuries after its explosion. Thus, if helium burning is followed by a relatively prompt detonation of the primary WD, the $^{44}$Ti gamma ray flux from a galactic or near-galactic SN Ia event will be detectable with Integral or similar MeV gamma ray missions. Such a strong $^{44}$Ti gamma may  distinguish any SN Ia originating from a DD merger, whether through a D6-like transient, or through a complete merger of the WD system in a classical double degenerate scenario,   from events from canonical near-$M_{\rm Ch}$ SNe Ia.\\

\begin {acknowledgements}
The authors commemorate the late E.G.B., who initiated this work with P.L.A., R.K., and R.T.F. several years ago. The authors acknowledge insightful conversations with R{\"u}diger Pakmor, Evan Bauer, and Noam Soker. R.T.F., N.R., and D.K. acknowledge support from NASA ATP award 80NSSC18K1013, NASA XMM-Newton award 80NSSC19K0601, and NASA HST-GO-15693. This work used the Extreme Science and Engineering Discovery Environment (XSEDE) Stampede 2 supercomputer at the University of Texas at Austin’s Texas Advanced Computing Center through allocation TG-AST100038. XSEDE is supported by National Science Foundation grant number ACI-1548562 \citep {townsetal14}.\\
\end {acknowledgements}

\software {FLASH 4.0.1 \citep{Fryxell_2000, dubeyetal12}, FLASH SN Ia module \citet{townsleyetal16} (\href{http://pages.astronomy.ua.edu/townsley/code}{http://pages.astronomy.ua.edu/townsley/code}), yt  \citep{Turk_2011}, Python programming language \citep{vanrossumetal1991}, Numpy \citep{vanderwaltetal2011}, IPython \citep{perezetal2007}, Matplotlib \citep{hunter2007}}.

\begin{singlespace}
	\bibliography{reference_d6.bib}
\end{singlespace}




\end{document}